\documentclass[prb,twocolumn,showpacs,amsmath,amssymb,superscriptaddress]{revtex4} 
\usepackage{bm}
\usepackage[]{graphicx}
\usepackage[tight]{units}
\usepackage{color}
\usepackage[colorlinks=true,citecolor=blue,linkcolor= blue,linkcolor=blue,urlcolor=blue]{hyperref}

\begin{document}
%
%
%
%
\title{Strong influence of the complex bandstructure on the tunneling
  electroresistance:\\ 
  A combined
  model and \textit{ab-initio} study}
\author{N. F. Hinsche}
\email{nicki.hinsche@physik.uni-halle.de}
\affiliation{Institut f\"{u}r Physik, Martin-Luther-Universit\"{a}t Halle-Wittenberg, D-06099 Halle, Germany}
\author{M. Fechner}
\affiliation{Max-Planck-Institut f\"{u}r Mikrostrukturphysik, Weinberg 2, D-06120 Halle, Germany}
\author{P. Bose}
\affiliation{Institut f\"{u}r Physik, Martin-Luther-Universit\"{a}t Halle-Wittenberg, D-06099 Halle, Germany}
\author{S. Ostanin}
\affiliation{Max-Planck-Institut f\"{u}r Mikrostrukturphysik, Weinberg 2, D-06120 Halle, Germany}
\author{J. Henk}
\affiliation{Max-Planck-Institut f\"{u}r Mikrostrukturphysik, Weinberg 2, D-06120 Halle, Germany}
\author{I. Mertig}
\affiliation{Institut f\"{u}r Physik, Martin-Luther-Universit\"{a}t Halle-Wittenberg, D-06099 Halle, Germany}
\affiliation{Max-Planck-Institut f\"{u}r Mikrostrukturphysik, Weinberg 2, D-06120 Halle, Germany}
\author{P. Zahn}
\affiliation{Institut f\"{u}r Physik, Martin-Luther-Universit\"{a}t Halle-Wittenberg, D-06099 Halle, Germany}

\date{\today}

\begin{abstract}
  The tunneling electroresistance (TER) for ferroelectric tunnel
  junctions (FTJs) with $\text{BaTiO}_{3}$ (BTO) and
  $\text{PbTiO}_{3}$ (PTO) barriers is calculated by combining the
  microscopic electronic structure of the barrier material with a
  macroscopic model for the electrostatic potential which is
  caused by the ferroelectric polarization. The TER ratio is
  investigated in dependence on the intrinsic polarization, the
  chemical potential, and the screening properties of the electrodes.
  A change of sign in the TER ratio is obtained for both barrier
  materials in dependence on the chemical potential. The inverse
  imaginary Fermi velocity describes the microscopic origin of
  this effect; it qualitatively reflects the variation and the
  sign reversal of the TER\@. The quantity of the imaginary Fermi velocity allows to obtain
  detailed information on the transport properties of FTJs by
  analyzing the complex bandstructure of the barrier material.
\end{abstract}
%
%
%
%
%
%

\maketitle

\section{\label{sec:level1} Introduction}
While in the last two decades the field of spintronics dominated the
area of high-technology memory and sensor devices, non-volatile ferroelectric memories gain more and more attention
\cite{Scott:1989p10155}. Recently, two phenomena which are
known in science for more than 80 years, namely
ferroelectricity and quantum mechanical tunneling, are combined in
ferroelectric tunnel junctions (FTJs) and could soon
compete with common spintronic devices \cite{Tsymbal2006}.

In one of the earliest concepts introduced by Esaki \textit{et al.}
\cite{Laibowitz:1971p10388}, that is, a `polar switch', the
tunneling barrier was composed of a ferroelectric. Its spontaneous, nonvolatile polarization can be switched by an 
applied electric field. At the best the switching of the 
barrier's polarization would create two different
electrostatic potentials and ,therefore, could lead to two distinct 
levels of the tunneling conductance. This tunneling 
electroresistance (TER) was predicted by Zhuravlev \textit{et al.}
\cite{Zhuravlev:2005p700,Zhuravlev2009} and could be employed in 
binary-logic devices.

While for a long time ferroelectricity was limited to dimensions above
several hundred nanometers and viewed as a collective phenomenon
\cite{Shaw2000}, its combination with quantum mechanical tunneling was
impossible. Hence, a TER effect could not be established.
However, during the past few years, there has been tremendous progress
in understanding finite-size effects in ultrathin ferroelectric films.
Today, theory predicts \cite{Junquera2003,Spaldin2004,Sai2005} and
experiment demonstrates \cite{Fong2004,Despont:2006p10177} the
presence of ferroelectricity for film thicknesses down to a few unit
cells; so, both support the concept of FTJs.  After the
breakthrough of Contreras \textit{et al.} \cite{Contreras2003}, a
number of recent experiments
\cite{Gajek2007,Garcia:2009p2133,Garcia:2010p8659,Crassous:2010p6876,Gruverman:2009p11002}
successfully investigated the transport properties of FTJs using
piezoresponse and conductive atomic force microscopy and found
evidence for a TER effect.

The purpose of this paper is
to investigate theoretically fundamental relations between the
electron transport through metal/ferroelectric/metal FTJ devices 
and the polarization-state-dependent electronic structure of the barrier. 

Here, we consider
both ultrathin ferroelectric $\text{BaTiO}_{3}$ (BTO) or
$\text{PbTiO}_{3}$ (PTO) as barriers.  Recently, calculations
based on free-electron models \cite{Zhuravlev:2005p700,Zhuravlev2009,Yang2007a,Zhuravlev:2009p2249} 
or first-principles methods \cite{Velev:2008p9558} explained the basics of the TER effect. 

In contrast to those investigations, our calculations successfully
combine an analytical and phenomenological model with
\textit{ab-initio} calculations.  This ansatz enables a systematic study of the TER in 
dependence on the barriers electronic structure and the position of the systems 
chemical potential $E_{F}$ separated from interface effects and the influence of 
the metallic electrodes.

For the electronic structure we
applied multiple-scattering theory, namely
Korringa-Kohn-Rostoker (KKR) methods
\cite{Zahn:1997p9722,Henk:2006p11004}, to obtain an accurate
state-of-the-art description of the electronic states (cf.\ Section~\ref{sec:estruct}). 
In all steps, we applied the local density approximation (LDA) to density-functional theory (DFT). The
calculation of the transport properties was realized within a
Landauer-B\"{u}ttiker picture \cite{Buttiker:1985p10457}, involving the
\textit{ab-initio} complex bandstructure and analytical expressions
for the electrostatic potential which is caused by the ferroelectric polarization and the 
different screening of the interface charges in the metallic electrodes
\cite{Zhuravlev:2005p700,Zhuravlev2009}. This approach provides a
flexible and fast computational means to observe the influence of 
material-specific parameters, especially the electrodes' properties, 
on the TER effect, while even taking into account the
exact electronic structure of the barrier. Therefor, the proposed
approach describes the transport quantities of FTJ's in a reliable way, which does 
not require a full self-consistent calculation of the whole junctions electronic structure.

The introduced term of the inverse imaginary Fermi velocity allows to estimate the 
TER effect for a given ferroelectric barrier material, knowing the barrier thickness, polarization strength and the the 
ratio of the electronic screening lengths in the electrodes.

This paper is organized as follows. In section \ref{sec:level2} the construction 
of the electrostatic potential in the concept of FTJ is introduced. Section \ref{sec:estruct} 
discusses issues of the self consistent electronic structure calculations, while 
section \ref{sec:tunnel} summarizes the calculation of the conductance. Section \ref{sec:results} presents the obtained results, 
starting with a detailed discussion of the complex bandstructure of BTO and PTO in \ref{sec:cbs} and focusing afterwards on the 
behavior of the TER in dependence of the position of the chemical potential (Sec.~\ref{sec:level4a}) and the electrodes electronic 
screening lengths (Sec.~\ref{sec:level4c}). The term of the inverse imaginary Fermi velocity is introduced in section \ref{sec:level4b} 
to provide a comprehensive relation between the electronic structure and the obtained transport properties.

\section{Methodology}
The electrostatic potential in  the barrier was derived from interface charges caused by the ferroelectric 
polarization of the material and the electronic screening in the electrodes. 
The electronic structure and so the decay rates of the states in the electronic 
band gap were calculated self-consistently. A perfect 2D periodicity was assumed perpendicular to the 
transport direction. The transmission probability was calculated using a WKB approximation. The matching 
of the wavefunctions at the interfaces was assumed to be equal for all states, so it does not enter 
the TER ratio in this approximation. Interface resonances which might rarely appear in real 
junctions are not considered and do not change the results qualitatively.

\subsection{\label{sec:level2} Ferroelectric tunnel barrier}
The physical mechanism behind the TER effect is the 
change of the electrostatic potential caused by the reversal of the
electric polarization $\bm{P}$ in the ferroelectric 
\cite{Zhuravlev:2005p700,Zhuravlev2009}. For sufficiently thin
ferroelectric films (a few unit cells thick), the charges
at the ferroelectric/electrode interfaces are not completely 
screened by the metallic electrodes. As a consequence, the
depolarizing electric field $\bm{E}$ is nonzero
\cite{Mehta1973,Junquera2003,Sai2005}, and its sign depends on
the direction of $\bm{P}$.

\begin{figure}[t]
\includegraphics[width=0.45\textwidth]{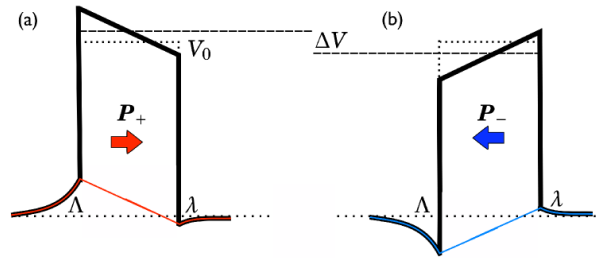}
\caption{\label{fig:1}(color online) Schematics of the tunnel barrier
  potential along the current direction for the ferroelectric
  insulator, with the barriers polarization pointing to the right (a,
  $P_{+}$) or to left (b, $P_{-}$) interface. The potential is a
  superposition of the electrostatic potential (colored line), the
  electronic potential which determines the bottom of the bands in the
  two electrodes, and the potential barrier created by the
  ferroelectric insulator. $\Lambda$ and $\lambda$ are the screening
  lengths of the electrodes, cf.\ Eq~(\ref{eq:phiz}).}
\end{figure}

Due to different screening properties of the electrodes, e.\,g.\
asymmetric interface termination \cite{Gerra2007,Velev:2008p9558} or
different electrode materials \cite{Contreras2003,Garcia:2010p8659},
the electric polarization leads to an asymmetry in the potential
profile.  Thus, the effective potential seen by electrons changes with
reversal of $\bm{P}$.  The averaged height of the barrier 
potential differs by $\Delta V$, which leads eventually to the TER effect (Fig.~\ref{fig:1}).

Assuming a Thomas-Fermi model of screening and a constant electric
field in the ferroelectric barrier, the static electric potential
profile in the electrodes and across the barrier of thickness $d$ can
be expressed as
\begin{equation}
  \varphi(z)=
  \frac{e \, \sigma_{S}}{\varepsilon_{0}}
  \begin{cases}
    \Lambda\,e^{-|z|/\Lambda} & z \le 0\\
    \Lambda - \frac{z}{d}(\lambda + \Lambda)
    & 0 \le z \le d\\
    -\lambda\,e^{-|z-d|/\lambda} & d \le z
  \end{cases},
  \label{eq:phiz}
\end{equation}
with $z$ the coordinate along the current direction
\cite{Zhuravlev:2005p700,Zhuravlev2009,Junquera2003}.  With the 
screening lengths $\Lambda$ and $\lambda$ of the left and the right
electrode, respectively, the screening charge per unit area is given
by $\sigma_{S}=\nicefrac{dP}{\epsilon (\Lambda+\lambda)+d}$, where $P$
is the absolute polarization and $\epsilon$ the relative static
permittivity of the barrier material. The latter was chosen for BTO
and PTO 1500 and 250, respectively \cite{Ghosez2006}. For simplicity,
$\Lambda$ is assumed to be larger than $\lambda$. The average height of the
potential barrier depends on the orientation of the polarization, 
this is, pointing to the left ($\bm{P}_{-}$) or to the right
($\bm{P}_{+}$) interface.  This is due to the fact that the additional
electrostatic potential has an averaged value of either 
$+\nicefrac{\left|\varphi(0)+\varphi(d)\right|}{2}$ or 
$-\nicefrac{\left|\varphi(0)+\varphi(d)\right|}{2}$.

\subsection{\label{sec:estruct}Electronic-structure calculations}
For our computational approach a sequential multi-code treatment
was applied. Various quantities were carefully cross-checked among the 
computer codes to obtain consistent results. Reliability is
achieved by numerous convergence tests.

Firstly, we determined the fully relaxed atomic positions according to 
Fechner \textit{et al.} \cite{Fechner2008a} within a pseudopotential
and plane-wave-basis scheme using \textsc{VASP} \cite{Kresse:1996p12346}. With the 
ferroelectric polarization pointing along the $+z$-direction, the
values of $\bm{P}$ for polar bulk BTO and PTO were calculated by a Berry-phase approach~\cite{Rabe2007}.

In a second step, the electronic structures of the considered 
systems, with atomic positions from the first step, were calculated 
self-consistently within the framework of density-functional theory 
(DFT), using a scalar-relativistic screened Korringa-Kohn-Rostoker 
(KKR) Green function method \cite{Zahn:1997p9722}. The spherical 
potentials were treated in the atomic sphere approximation (ASA), 
using the local density approximation (LDA) for the 
exchange-correlation potential \cite{Hohenberg1964,Vosko1980}.

In a third step, the complex band structure $k_{z}(E,\bm{k_{\parallel}})$ 
was computed within a layer KKR, using the same potentials as in the second stage \cite{Henk:2006p11004, Henk2001}. 
The imaginary part of $k_{z}(E,\bm{k_{\parallel}})$ will be denoted as decay parameter $\kappa(E,\bm{k_{\parallel}})$.
As will become clear in what follows (\ref{sec:cbs}), evanescent 
states play an important role in the tunneling regime.

Applying the scheme explained in the upcoming subsection, we are now
able to compute the transmission probability which determines the
conductance and therefor the TER properties.

\subsection{\label{sec:tunnel}Conductance calculations}
In order to predict the conductance change associated with a
polarization switching, we assume so thin a barrier that the
dominant transport mechanism is quantum-mechanical tunneling and the
ferroelectricity is preserved \cite{Fong2004,Spaldin2004}.
Therefore, four unitcells ($d \approx$ \unit[1.6]{nm}) of BTO or PTO
are considered as electric switchable barrier.

Assuming elastic and coherent transport, the zero-bias 
conductance per unit cell area is given by the 
Landauer-B\"{u}ttiker formula \cite{Buttiker:1985p10457}.

\begin{equation}
  G = \frac{2e^2}{h}\sum\limits_{\bm{k_{\|}}}T(E_F,\bm{k_{\|}}),
\label{for:landauer}
\end{equation}
where
\begin{equation}
  T(E_F,\bm{k_{\|}})
  =
  T_{0}\,\exp[-2\int\limits_{0}^{d}dz\, \kappa(E_{F}-\varphi(z),\bm{k_{\|}})]
\label{for:trans}
\end{equation}
is the transmission probability in the WKB approximation at the
chemical potential $E_{F}$. $\kappa$ is the smallest imaginary part of the complex wavevector (cf.\ Section~\ref{sec:cbs}) and is calculated by means of 
the \textit{ab-initio} complex band structure. The WKB 
approximation takes into account the shape of the barrier potential 
$\varphi(z)$ and the decay properties of the electronic states in the 
barrier gap. The $\bm{k_{\|}}$ integration is over the whole 
two-dimensional Brillouin zone, using at least 6000 special
points in the irreducible part of the Brillouin zone.
The transmission prefactor $T_{0}$ comprises the influence of the wavefunction matching 
at the interfaces to the tunneling probability. We used a quite rough approximation, taken $T_{0}$ 
to be constant for all energies and inplane wavevectors $\bm{k_{\|}}$. This factor drops out in the 
determination of the TER according to Eq.~(\ref{for:ter}). The change of the potential in the 
electrodes in the vicinity of the interfaces is not included, because metals with a constant 
density of states are assumed.

A main feature of transport through FTJs is a conductance 
asymmetry, defined as
\begin{equation}
  \text{TER} 
  = 
  \frac{G(\bm{P}_{-})-G(\bm{P}_{+})}{G(\bm{P}_{-})+G(\bm{P}_{+})},
  \label{for:ter}
\end{equation}
where $G(\bm{P}_{+})$ ($G(\bm{P}_{-})$) is the conductance for
the polarization pointing to the right (left) interface.

\section{\label{sec:results}Results and discussion}
\subsection{\label{sec:cbs}Complex band structures}
To prepare material-specific input for the transport calculations, it
is useful to recall the importance of the complex band structure 
for tunneling through an (ferroelectric) insulator (see 
\onlinecite{Mavropoulos2000} for magnetic tunnel junctions). The imaginary 
part $\kappa$ of the complex wavevector dominates the transport properties; 
its product with the barrier thickness defines the exponential decay 
of the wavefunctions within the barrier and determines essentially
the transmission probability.

In a periodic crystal, the wavevectors $\bm{k}$ of Bloch states 
are necessarily real. Heine~\cite{Heine1965} demonstrated that a 
surface or interface state can be obtained by matching 
eigenstates of both sides of the boundary plane. This procedure 
requires in addition states with an imaginary or---more
general---a complex wave vector, i.\,e.\ so-called evanescent Bloch
states. The wave-function matching across the boundary requires
eigenstates with complex wave numbers $k_{z}$ at a given real energy
have to be taken into account.

At an interface between a metal and an insulator, prominent states are
the metal-induced gap states (MIGS), which are itinerant in the metal
electrodes but decay exponentially into the insulating barrier. The
dispersion relation of the evanescent states is the so-called complex
band structure.  For a planar tunnel junction, the periodicity along
the interfaces (barrier) requires that the projection $\bm{k_{\|}}$ of
the wave vector $\bm{k}$ onto the interface is real and conserved.
However, the perpendicular component $k_{z}+ i \kappa$ can be complex
and depends on the region (electrodes, barrier).

\begin{figure}
  \includegraphics[width=0.48\textwidth]{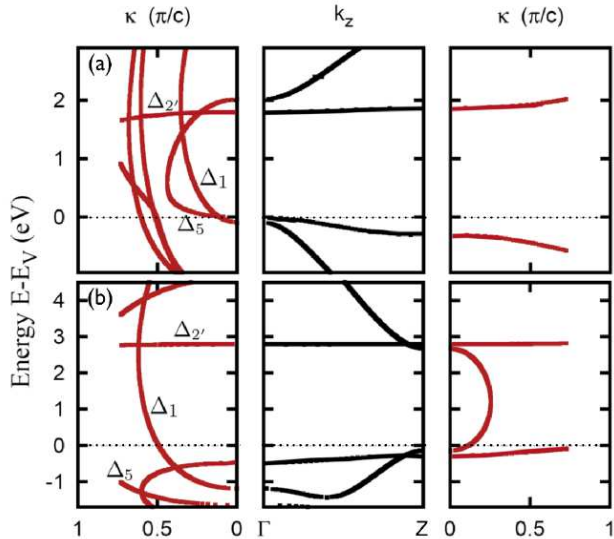}
  \caption{\label{fig:2}(color online) Symmetry-resolved complex band
    structures for ferroelectric (a) $\text{BaTiO}_{3}$ and (b)
    $\text{PbTiO}_{3}$ at $\overline{\Gamma}$ in the 2D Brillouin
    zone ($\bm{k_{\|}}=0$). Middle panels show the conventional
      bands structure ($\operatorname{Im}\,k_{z}=0$) along the
      $\Gamma-Z$-line. Left and right panels display imaginary bands
    of the first ($k_{z}=0$) and the second kind
    ($k_{z}=\frac{\pi}{c}$), respectively.}
\end{figure}

From Figure \ref{fig:2} we deduce fundamental band gaps of 
ferroelectric BTO and PTO of \unit[1.79]{eV} and \unit[2.89]{eV},
respectively.  Both are increased, by $4.1\%$ for BTO and $21.4\%$ for
PTO, as compared to the calculated paraelectric case.  The strong localization of 
the $\Delta_{2^{\prime}}$ states prevents larger band gaps.

The para-to-ferroelectric phase transition lowers the symmetry 
from cubic (Pm3m) to tetragonal (P4mm). At $\overline{\Gamma}$, the
bands can be decomposed with respect to the irreducible representation
of the point group 4mm, as shown in Fig.~\ref{fig:2}. The 
$t_{2g}$ bands split to form a doubly degenerate $\Delta_5$ band
($d_{zx}$, $d_{zy}$) and a $\Delta_2$ band ($d_{xy}$). Similarly, the 
$e_g$ bands split into a $\Delta_1$ band ($d_{z^2}$) and a 
$\Delta_{2^{\prime}}$ band ($d_{x^{2}-y^{2}}$).

According to Chang \cite{Chang:1982p710}, complex bands can be 
classified as follows. \textit{(i)} Real bands correspond to the
conventional band structure and have $\operatorname{Im}\,k_{z}=0$.
Thus, the wave functions are the Bloch states. \textit{(ii)}
Imaginary bands of the first kind have $\operatorname{Re}\,k_{z}=0$
and $\operatorname{Im}\,k_{z} \ne 0$. \textit{(iii)} Imaginary bands
of the second kind have $\operatorname{Re}\,k_{z}=\nicefrac{\pi}{c}$
and $\operatorname{Im}\,k_{z} \ne 0$. \textit{(iv)} Complex bands have
$\operatorname{Re}\,k_{z} \ne 0$ , $\operatorname{Re}\,k_{z} \ne
\nicefrac{\pi}{c}$ and $\operatorname{Im}\,k_{z} \ne 0$ with 
c being the periodicity in z-direction.

At $\overline{\Gamma}$, the three states with the smallest decay rates
within the fundamental band gap consist of a $\Delta_5$ doublet and a
$\Delta_1$ singlet for both perovskite ferroelectrics. Neverthless, we 
address, that the smallest decay rate for BTO is formed by either
the $\Delta_5$ or the $\Delta_1$ imaginary band of first kind
(cf.~\ref{fig:2}(a), band crossing in the left panel), while for PTO
the smallest decay parameter is always associated with a 
$\Delta_1$-like imaginary band of the second kind 
(cf.~\ref{fig:2}(b), right panel). Importantly, these complex
bands---and hence the electric conductance---are highly
sensitive to the ferroelectric displacements; the latter, in turn,
depend significantly on the polarization direction and on the
potential profile in the barrier.

\begin{figure}
\includegraphics[width=0.48\textwidth]{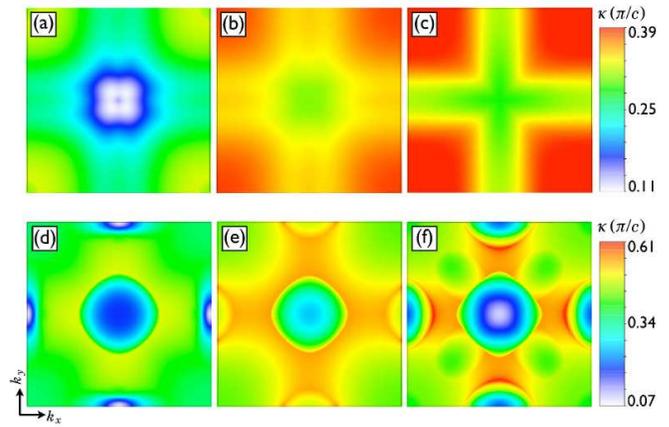}
\caption{\label{fig:3}(color online) Smallest imaginary part of the
  complex wavevector of $\text{BaTiO}_{3}$ (upper row, a--c) and
  $\text{PbTiO}_{3}$ (lower row, d--f) in the 2D Brillouin zone (edges
  of the Brillouin zone at $\pm\nicefrac{\pi}{a}$). The imaginary
  parts are shown for different energies relative to the valence band
  maximum: (a) $0.77~\mbox{eV}$, (b) $1.30~\mbox{eV}$, (c)
  $1.77~\mbox{eV}$, (d) $1.10~\mbox{eV}$, (e) $1.40~\mbox{eV}$, (f)
  $2.10~\mbox{eV}$.}
\end{figure}

Since at each $\vec{k}_{\parallel}$ the transmittance is determined by
the least decaying wavefunction, we address the $\bm{k_{\|}}$-resolved
smallest imaginary part $\kappa$ of the wave vector at different
energies in Fig.~\ref{fig:3}. For BTO, the main contribution 
to the transmission is expected to come from a small annulus around 
$\overline{\Gamma}$ for all energies within the gap.  This is caused 
by the smallest decay rates near the Brillouin zone center. Our 
findings corroborate those of Velev \textit{et
  al.}~\cite{Velev:2007p665}. In contrast to the paraelectric phase
(not shown), the smallest imaginary part of the wavevector is not
located exactly at the Brillouin zone center but slightly off-set for
energies close to the valence band edge (Fig.~\ref{fig:3}(a)). This is
explained by the lifted degeneracy of the $\Delta_1$ and $\Delta_5$
bands at the $\overline{\Gamma}$-point. The latter supports the
importance of a wave-vector analysis in the entire Brillouin
zone.

For PTO, the situation is more complex in comparison to
BTO\@. In particular, the smallest $\kappa$ are located at
$\overline{\Gamma}$ and $\overline{X}$ {edge centers of the 2BZ 
for energies close to the valence band maximum and the conduction band
minimum (\ref{fig:3}(d)-(f)). In the middle of the gap, states from
the Brillouin zone center dominate the transport
(\ref{fig:3}(e)). In contrast to BTO, one depicts for PTO nearly the same
damping around $\overline{M}$ (corners of the Brillouin zone), while
sizably larger $\kappa$ in the rest of the Brillouin zone can be stated (note the different scale in top and bottom panels of Fig.~\ref{fig:3}).
Thus, one expects remarkably different $\bm{k_{\|}}$-transmissions 
and smaller tunneling currents as compared to BTO\@.

\subsection{\label{sec:level4}Transport Properties}
Having provided complex band structures, the tunneling
conductance of the FTJs is investigated in dependence on the
polarization of the barrier, the screening lengths in the 
electrodes, and the chemical potential with respect to the fundamental
band gap of the barrier. For ferroelectrics, the Fermi energy can be
adjusted within the gap by crystal doping \cite{Schwartz2000}, while a
shift of the chemical potential can be realized by a small bias.

\subsubsection{\label{sec:level4a}Energy dependence of the TER}

We recall that the slope of the imaginary part of the complex
wavevector $\kappa$ and the $\kappa$ distribution in the Brillouin
zone change remarkably with energy (cf.\ Figs.~\ref{fig:2} and
\ref{fig:3}). As a consequence, we focus on three positions in 
the gap: close to the valence band, in the center of the gap and close
to the conduction band.

\begin{figure*}
\includegraphics[width=0.75\textwidth]{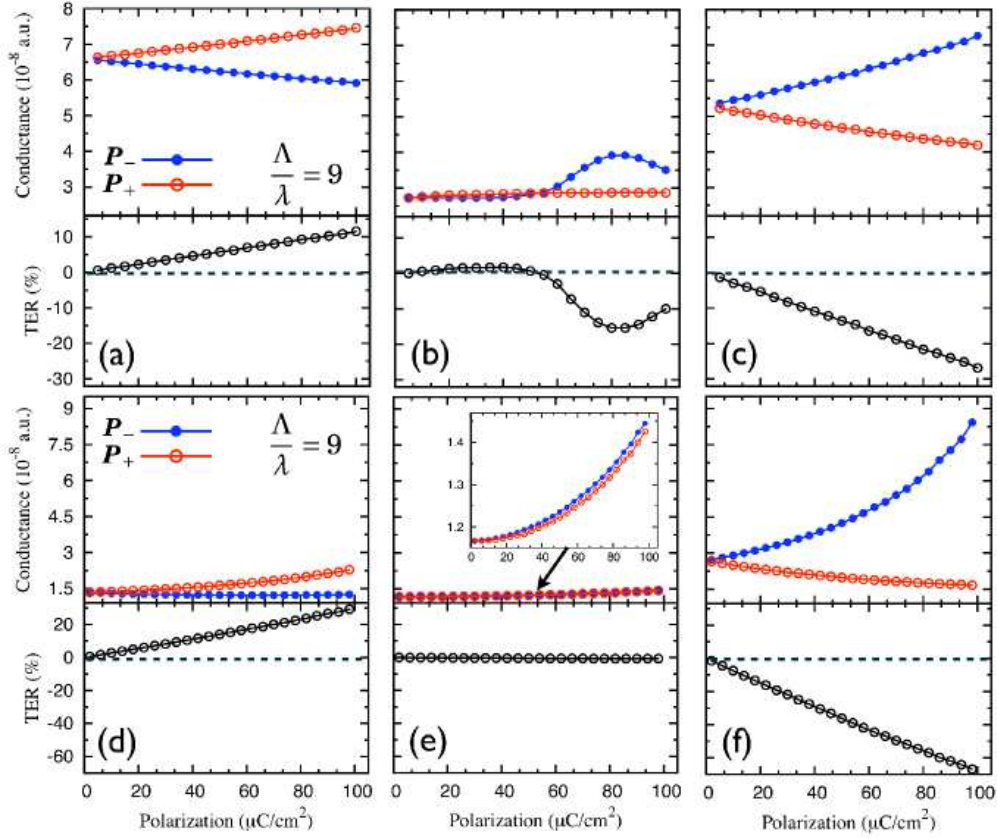}
\caption{\label{fig:4}(color online) Conductances and tunneling
    electroresistances (TERs) for the two ferroelectric
  configurations of a FTJ with $\text{BaTiO}_{3}$ (upper row, a--c)
  and $\text{PbTiO}_{3}$ (lower row, d--f) as a ferroelectric
  barrier. The barrier thickness was fixed to four unit cells
  ($\approx$ \unit[1.6]{nm}).  Both, conductance and TER, are shown in
  dependence on the absolute value of the electrical polarization
  as well as for different chemical potentials with respect to the
  valence band maximum: (a) $0.77~\mbox{eV}$, (b) $1.30~\mbox{eV}$,
  (c) $1.77~\mbox{eV}$, (d) $1.10~\mbox{eV}$, (e) $1.40~\mbox{eV}$,
  (f) $2.10~\mbox{eV}$ (that is, as in Fig.~\ref{fig:3}). Full
  blue (open red) circles denote conductances for the
  polarization pointing to the left (right) interface. The inset in
  (e) depicts the conductances in a smaller scale, showing a
  small but pronounced variation.  $\Lambda$ and $\lambda$ are
  the screening lengths in the left and right electrode,
  respectively , with fixed ratio $\nicefrac{\Lambda}{\lambda} = 9$.}
\end{figure*}

In Figure \ref{fig:4} the calculated conductance and TER for
FTJ's with BTO (upper row, a–-c) and PTO (lower row, d-–f ) as a
ferroelectric barrier material are presented.  In the upper part of
each panel the conductance referring to Eq.~(\ref{for:landauer}) is
shown. Full blue (open red) circles
denote the dependence of the conductance
for the polarization pointing to the left (right) interface, while the ratio of 
screening lengths of the electrodes are fixed at 
$\nicefrac{\Lambda}{\lambda} = 9$. This might refer to a noble
metal electrode ($\lambda_\text{Ag}=\unit[0.07]{nm}$) and a
ferromagnetic electrode ($\Lambda_{\text{SrRuO}_3}=\unit[0.6]{nm}$) or
a half metallic electrode
($\Lambda_{\text{La}_{1-x}\text{Sr}_{x}\text{MnO}_{3}}=\unit[0.2-1.9]{nm}$)
as reported in \cite{Junquera2003,Zhuravlev:2005p700,Gajek2007}. In
the lower parts of the panels \ref{fig:4} (a)-(f) the TER determined
by Eq.~(\ref{for:ter}) in dependence on the abolute value of the
polarization is represented by open black circles.

As we consider different energies in the gap---close to the
valence band (a,d), near the gap center (b,e), and close to the
conduction band (c,f)---one easily recognizes the increase of the
conductance near the gap edges. This is readily explained by the 
imaginary part of the wavevectors which decreases for certain
$\bm{k_{\|}}$ (compare Figs.~\ref{fig:2} and \ref{fig:3}). Hence, 
the transmission probability and the conductance
(ref.\ to Eq.~(\ref{for:trans})) increase. In a free electron 
picture as used in previous explanations of the TER effect
\cite{Zhuravlev:2005p700,Zhuravlev2009,Yang2007a,Zhuravlev:2009p2249},
the conductance would decrease continuously by shifting the chemical
potential from the band edges. Our approach takes into account a more realistic 
electronic structure of the ferroelectric barrier.

The behavior of the conductance for different chemical potentials can
be understood qualitatively by the complex band structure at
$\bar{\Gamma}$ (see Fig.~\ref{fig:2}).  For energies near
the valence band the $\bm{P}_{+}$-conductance is larger than 
the $\bm{P}_{-}$ conductance (compare Fig.~\ref{fig:4}(a,c)). This is caused
by the lower decay rates for the electron wave functions contributing
to the tunneling current in the $\bm{P}_{+}$ state. The shift of the bands 
to higher energies due to the positive contribution of the
electrostatic potential leads to smaller values of the imaginary part
of the wave vector at the chemical potential. Taking into account the
definition of the TER (ref. Eq.~(\ref{for:ter}) a positive TER occurs.
For energies close to the conduction band edge the behavior of the
conductances and, thus, of the TER reverse, as does the curvature 
of the imaginary bands reverse, too.

For bulk polarizations of $\bm{P}_{\mathrm{BTO}} = \unit[22.9]{\mu
    C}/\unit{cm^{2}}$ and $\bm{P}_{\mathrm{PTO}} = \unit[94.3]{\mu
    C}/\unit{cm^{2}}$ \cite{Fechner2008a} we achieve high TER values
of nearly \unit[3]{\%} ($E_F=E_V+\unit[0.77]{eV}$) and \unit[27]{\%}
($E_F=E_V+\unit[1.10]{eV}$) for chemical potentials near to the
valence band and allmost \unit[6]{\%} ($E_F=E_V+\unit[1.77]{eV}$) and
\unit[61]{\%} ($E_F=E_V+\unit[2.10]{eV}$) for energies close to the
conduction band. 
The value for PTO is in agreement with
  experimental values of nearly \unit[60]{\%} for a FTJ with
Pb(Zr,Ti)O as ferroelectric barrier with different electrode materials
\cite{Contreras2003}. Recent findings for FTJs based on strained
BTO show a TER of \unit[16]{\%} \cite{Garcia:2010p8659}, which is in
astonishing agreement with our predictions (see
Fig.~\ref{fig:4}(c)) for strained thin films (note that
then $\bm{P}_{\mathrm{BTO}} = \unit[50]{\mu C}/\unit[]{cm^{2}}$
according to \cite{Ederer2005}).

As for energies lying close to the gap edges the discussion can be
made in a generally more intelligible picture, the situation is quite
more complex for energies in the middle of the gap. In figure
\ref{fig:4}(b) the evaluation of the TER for a FTJ consisting of BTO
is shown for $E_F=E_V+\unit[1.30]{eV}$ near the point where the
symmetry character of the dominating evanescent wave functions at the
$\Gamma$-point changes from $\Delta_1$ to $\Delta_5$ (compare
Fig.~\ref{fig:2}(a), left panel) and consequently the slope of 
the imaginary bands change sign.
  
For polarization strengths less than $\bm{P}_{\mathrm{BTO}} = \unit[50]{\mu C}/\unit[]{cm^{2}}$ 
the TER nearly vanishes, while for higher
polarization strengths the conductance for the $\bm{P}_{-}$ state
increases and a negative TER of $\unit[-15]{\%}$ can be reached.
As noted earlier, those higher polarization strengths might be
achieved in stressed thin films where values of
$\bm{P}_{\mathrm{BTO}} =\unit[60]{\mu C}/\unit[]{cm^{2}}$ and
even $\bm{P}_{\mathrm{PTO}} = \unit[110]{\mu C}/\unit[]{cm^{2}}$ 
were reported \cite{Ederer2005}. In the case of PTO the TER is barely
visible for the whole range of polarizations with values below
\unit[1]{\%} for an energy of $E_F=E_V+\unit[1.40]{eV}$. This is near
the center of the complex wave vectors loop as shown in
Fig.~\ref{fig:2}(b), right panel.  Close to the maximum of the
band with respect to the decay rates, a small shift of the potential
caused by the electric polarization reversal causes a negligible
change of the decay rates, only.

As the decay parameters of all states in the Brillouin zone are
comprised for the calculation of the transport properties, different
curvatures of $\kappa$ are involved in the total resulting TER and a
straightforward explanation of the properties of the TER is hardly 
possible. In the preceding, we established major trends by discussing three 
exemplary energies. Now, we turn to the TER in dependence on the
position of the chemical potential relative to the barrier gap. In 
the upper row of Fig.~\ref{fig:5}, the most remarkable feature is a
change of sign of the TER at energies of about \unit[1.5]{eV} for both
materials. This result cannot be explained by a free electron
model, where the negative slope of the imaginary band would lead to
negative values of the TER over the whole range of the band gap.
Furthermore, nearly free electron models or tight-binding
approaches would produce only one sign change at an energy
approximately in the middle of the gap, which is at about \unit[0.8]{eV} and
\unit[1.4]{eV} for BTO and PTO, respectively.

For the chosen polarization values and for energies close to the
valence band of BTO, the TER is expected to be as twice as large as
for energies close to the conduction band. For PTO even higher values of
\unit[40]{\%} (close to the valence band) and \unit[-60]{\%} (close
to the conduction band) are expected.

\begin{figure}
\includegraphics[width=0.48\textwidth]{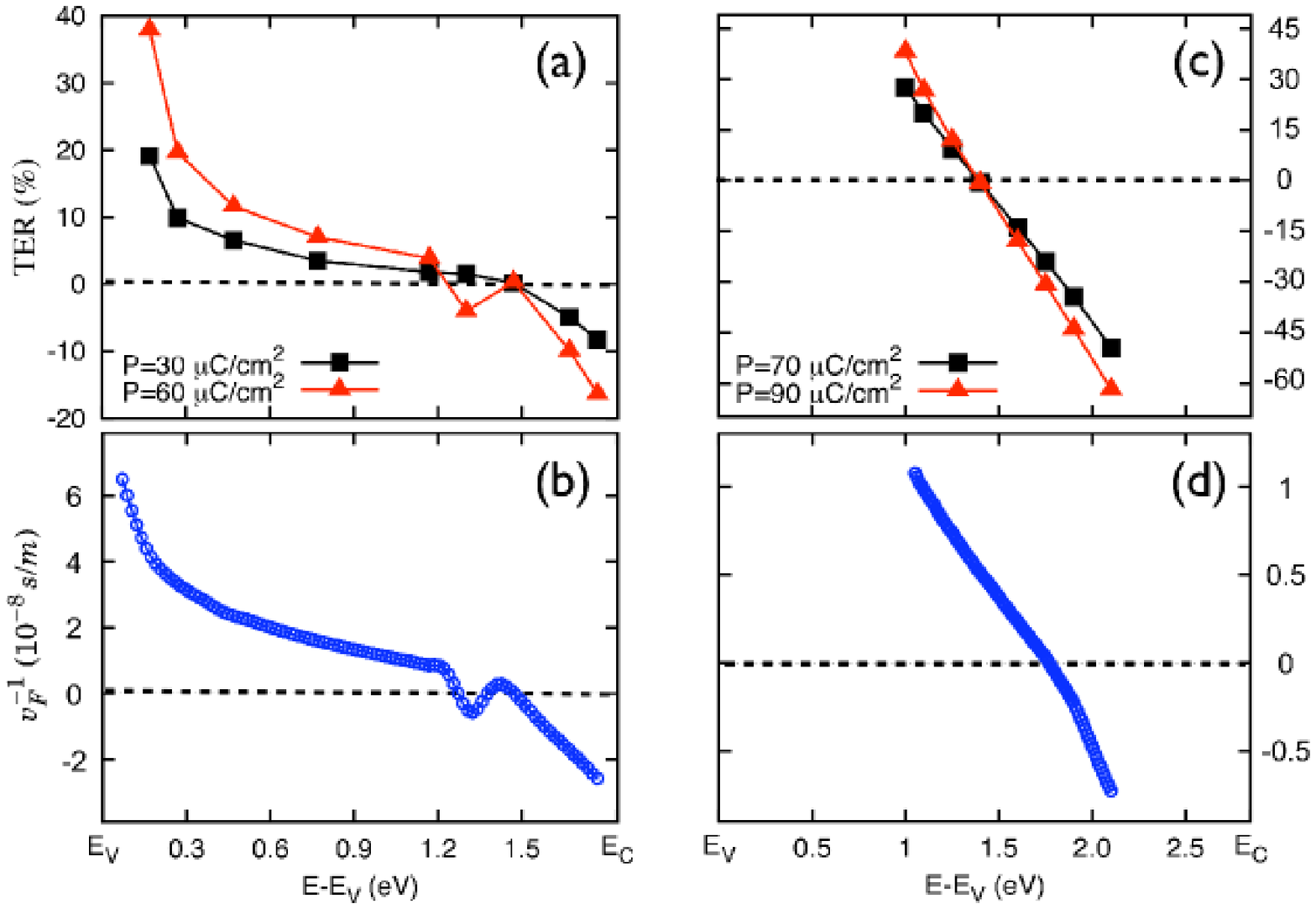}
\caption{\label{fig:5}(color online) Tunneling electroresistance
    (TER) versus chemical potential for BTO (a) and PTO (b) relative
    to the valence band maximum of the ferroelectric. The screening
  lengths' ratio is fixed to 9.  In the lower part the
  corresponding inverse imaginary Fermi velocity for BTO (c) and PTO
  (d) is depicted. For further details see text.}
\end{figure}

\subsubsection{\label{sec:level4b} Inverse imaginary Fermi Velocity}

A qualitative explanation for the previous described energy dependent behavior of the TER 
is provided by the introduction of the inverse imaginary Fermi velocity 

\begin{equation}
  \nu_{F}^{-1}(E)
  =\hbar \,
  \langle\frac{d\kappa(\bm{k_{\|}},E)}{dE}\rangle_{\mathrm{BZ}}.
\end{equation}

This quantity can be seen as the
$\bm{k_{\|}}$-averaged energy derivative of the decay parameter
$\kappa$ of the complex band structure in the Brillouin zone. If this
parameter is small, the additional electrostatic potential caused by
the electrical polarization $\bm{P}_{-}$ or $\bm{P}_{+}$ does not
change the effective decay parameter remarkably. So, a small TER
effect is expected. On the contrary, a large energy derivative of
the decay parameter $\kappa$ which most likely occurs close to the 
band edges points to a large TER effect.

To illustrate the correlation of the inverse imaginary Fermi velocity
and the TER, an approximation for a single $\bm{k_{\|}}$ is
given in the following. Small polarizations will lead to small
  differences in the complex bandstructure for both $P_{\pm}$. Hence, 
  the imaginary part $\kappa_{-}$ of the complex wavevector $k_z$ for
$P_{-}$ can be written as
\begin{equation}
  \kappa_{-}
  =
  \kappa_{+} + \frac{\partial \kappa}{\partial E} \cdot \Delta V,
  \label{for:dk}
\end{equation}
by expanding the imaginary part $\kappa_{+}$ for $P_{+}$.
The TER given in Eq.~(\ref{for:ter}), including Eqs.~(\ref{for:trans}) and (\ref{for:dk}) is directly heading
to the approximation
\begin{equation}
  \text{TER}\approx \alpha \cdot (\hbar \cdot \nu_{F})^{-1} \cdot P \cdot d^{2}.
\label{for:appter}
\end{equation}
Here,
\begin{equation}
  \alpha=\frac{e \, (\Lambda-\lambda)}{\epsilon_0 \, [\epsilon \,(\Lambda+\lambda)+d]},
\end{equation}
is a unique constant for each junction, as $\epsilon$ and $d$
contain material-specific information of the ferroelectric
barrier and the screening lengths $\Lambda$ and $\lambda$ are
determined by the electrodes.  An explicit value of $\alpha$ for BTO
($\Lambda = \unit[6]{nm}$, $\lambda = \unit[0.7]{nm}$, $\epsilon =
  1500$, $d = \unit[1.6]{nm}$) would be
$\alpha\approx\unit[60 \cdot 10^{6}]{e Vm}/\unit[]{C}$.

The effect of all states in the whole Brillouin zone has to be
recognized to describe the TER effect in a reliable way. Therefor the
inverse imaginary Fermi velocity is a combined quantity for the
damping of all wavefunctions in the Brillouin zone and hence a value
for the probability of transmission.  The lower panels in Fig.\ref{fig:5} 
depict the inverse imaginary Fermi velocity for
ferroelectric BTO (b) and PTO (d). The behavior of the TER is
fully reproduced qualitatively and quantitatively. To be more
  specific, the energy position of the sign change in the TER as well
  as the magnitude of the TER for different chemical potentials can
  be achieved by analyzing the inverse imaginary Fermi velocity. The
different ratios between the TER and the inverse imaginary Fermi
velocity are caused by the relative static permittivities
for BTO and PTO\@. This also prevents the calculation of the TER for
PTO junctions for chemical potentials close to the band edges, as
electrical breakthrough and metallic behavior would be the consequences.

The perfect agreement of the behavior  of TER and the inverse imaginary Fermi velocity is to some extent accidental. 
The average over $\nu_{F}^{-1}$ is taken in the whole 2BZ, whereas the transmission defined in 
Eq.~(\ref{for:trans}) is dominated by the $\bm{k_{\|}}$ states with the lowest decay parameters 
$\kappa(E,\bm{k_{\|}})$. It seems that these regions of the 2BZ are also dominated by the averaged 
inverse Fermi velocity. For the barrier materials BTO and PTO the latter could be found. Introducing the exact material properties of the electrodes, the symmetry selection 
caused by the wavefunction matching conditions at the interfaces would most likely weaken the link between the TER and $\nu_{F}^{-1}$.

\begin{figure}
  \includegraphics[width=0.48\textwidth]{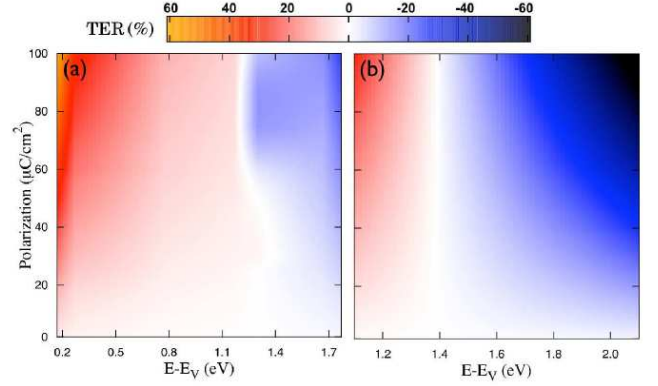}
  \caption{\label{fig:6}(color online) TER dependence on the
    polarization strength and the position of the chemical potential
    for FTJ with (a) BTO or (b) PTO barriers. The screening
    lengths' ratio was fixed to $\nicefrac{\Lambda}{\lambda}=9$.}
\end{figure}

To complete the discussion, we show in Fig.~\ref{fig:6} the
dependence of the TER on $\bm{P}$ and $E_{\mathrm{F}}$. For BTO,
the energetic position of the sign change of the TER varies slightly
with the polarization strength, which is most likely linked to the
change of symmetry character of the dominating evanescent wave
functions associated with a sign reversal of the inverse imaginary Fermi velocity 
at the crossing point of the two imaginary bands of the first kind 
at an energy of about \unit[1.4]{eV} above the valence band edge 
(compare top panel in Fig.\ref{fig:2}). In the case of PTO
the sign change is obtained for a chemical potential of about
\unit[1.4]{eV} above the valence band edge, independent on the
intrinsic polarization strength. Again it is conspicuous 
that the highest values for the TER occur near the band edges, where
the inverse imaginary Fermi velocity is maximized. The largest
absolute values for the TER are found near the valence band
maximum or conduction band minimum for BTO or PTO, respectively.

\subsubsection{\label{sec:level4c}Influence of the electrodes' screening lenghts}

To get more insight on how the TER depends on the electrodes'
  properties, we present in Fig.~\ref{fig:7} its dependency on the
  ratio of the screening lengths for $\bm{P}_{\mathrm{PTO}} =
  \unit[70]{\mu C}/\unit[]{cm^{2}}$. The spreading of the
conductances for $\bm{P}_{-}$ and $\bm{P}_{+}$ increases with the
difference of the screening lengths. An interchange of the
electrodes will not affect the conductances, as only the
  polarization direction is reversed. An asymmetry of the screening
lengths in the electrodes is mandatory to obtain a nonzero TER\@.
For equal screening lengths, one obtains $G(\bm{P}_{-}) = G(\bm{P}_{+})$ and,
  hence, the TER vanishes. On the other hand, an interchange of the
screening lengths in the electrodes reverses the sign of the TER as a
result of the definition in Eq.~(\ref{for:ter}).  If one interface is
a free surface, the ratio of the screening lengths will be maximized,
leading to giant values of the TER as observed with conductive atomic
force microscopy on BTO and PTO surfaces
\cite{Garcia:2009p2133,Gruverman:2009p11002,Crassous:2010p6876}.

\begin{figure}
  \includegraphics[width=0.48\textwidth]{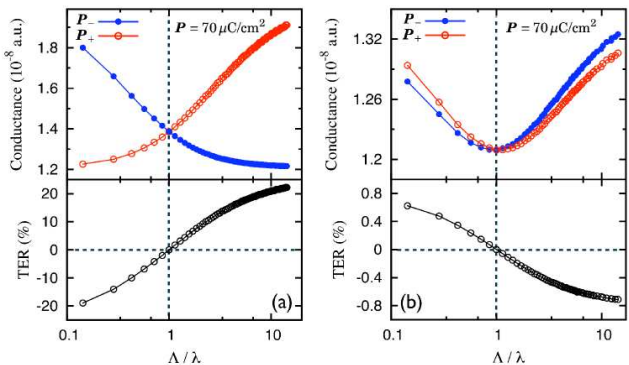}
  \caption{\label{fig:7}(color online) The conductance for the two
    ferroelectric polarization states and the resulting TER of a FTJ
    with $\text{PbTiO}_{3}$ as a ferroelectric barrier material is
    shown. Full blue (open red) circles denote the dependence of the
    conductance for the polarization pointing to the left (right)
    interface. The absolute polarization is fixed to
    $\bm{P}_{\text{PTO}}=\unit[70]{\mu C}/\unit[]{cm^{2}}$. Both,
    conductance and TER, are shown in dependence on the ratio $\nicefrac{\Lambda}{\lambda}$ of the
    screening lengths of the electrodes material and for different
    chemical potentials with respect to the valence band: (a)
    $1.10~\mbox{eV}$, (b) $1.40~\mbox{eV}$.}
\end{figure}

\section{\label{sec:level5}Conclusion and outlook}
The tunneling electroresistance in ferroelectric tunnel junctions 
with BTO and PTO barriers was discussed in dependence on the position
of the chemical potential, the electrical polarization, and the
ratio of the screening lengths in the electrodes. In dependence
on the chemical potential position a sign reversal of the
TER ratio is obtained. This is caused by a sign change of the
  energy derivative of the decay parameter $\kappa$. The latter was
  comprised by the inverse imaginary Fermi velocity. The behavior of
this quantity in the band gap is in agreement with the TER
ratios.

As a consequence of the presented results we demonstrate that the
analysis of the complex band structure is a powerful tool to gain
useful insight into the behavior of the TER effect more precise than the analysis 
of free electron like models and more efficient than calculations of the full contact geometry done in previous works.

As a sign change of the TER in dependence on the chemical potential is achieved, an experimental proof by application of different electrode materials 
or doping levels is proposed. We again emphasize that the shift of the chemical potential and thus the desired sign change of the TER might be also achieved 
by a small applied bias.

\begin{acknowledgments}
  This work was supported by the Deutsche
  Forschungsgemeinschaft, SFB 762 `Functionality of Oxide
    Interfaces'. N. F. Hinsche, M.  Fechner, and P. Bose are
  members of the International Max Planck Research School for Science
  and Technology of Nanostructures.
\end{acknowledgments} 

\bibliography{paperJH1_NH.bbl}
\end{document}